\documentclass[11pt,a4paper]{article}

\usepackage{amssymb,amsmath}

\newcommand{\bb}{\begin{equation}}
\newcommand{\ee}{\end{equation}}
\newcommand{\Z}{{\mathbb Z}}
\newcommand{\N}{{\mathbb N}}
\newcommand{\ICG}{\mathrm{ICG}}
\renewcommand{\i}{\mathrm i}
\newcommand{\R}{{\mathbb R}}

 \newtheorem{thm}{Theorem}
 \newtheorem{prop}[thm]{Proposition}
 \newtheorem{lem}[thm]{Lemma}
 \newtheorem{cor}[thm]{Corollary}

\newcommand{\QED} {\hfill$\square$}

\newenvironment{dok} {\par \noindent \textbf{Proof. }}{\QED \par \bigskip \par}

   \font\sst=cmtt8
  \font\ssi=cmti8 
  \font\sst=cmtt10 
 \font\ssi=cmti10

\title{\bf Characterization of circulant graphs having perfect state transfer}

\begin{footnotesize}
\author{\frenchspacing
Milan Ba\v si\'c\\
{\ssi Faculty of Sciences and Mathematics, University of Ni\v{s},}\\
{\ssi Vi\v segradska 33, 18000 Ni\v s, Serbia} \\
{\ssi E-mail:} {\sst basic\_milan@yahoo.com} \\
}
\end{footnotesize}

\date{}

\begin{document}

\maketitle

\begin{abstract}
In this paper we answer the question of when circulant quantum spin
networks with nearest-neighbor couplings can give perfect state
transfer. The network is described by a circulant graph $G$, which
is characterized by its circulant adjacency matrix $A$. Formally, we
say that there exists a {\it perfect state transfer} (PST) between
vertices $a,b\in V(G)$ if $|F(\tau)_{ab}|=1$, for some positive real
number $\tau$, where $F(t)=\exp(\i At)$. Saxena, Severini and
Shparlinski ({\it International Journal of Quantum Information} 5
(2007), 417--430) proved that $|F(\tau)_{aa}|=1$ for some $a\in
V(G)$ and $\tau\in \R^+$ if and only if all eigenvalues of $G$ are
integer (that is, the graph is integral). The integral circulant
graph $\ICG_n (D)$ has the vertex set $Z_n = \{0, 1, 2, \ldots, n -
1\}$ and vertices $a$ and $b$ are adjacent if $\gcd(a-b,n)\in D$,
where $D \subseteq \{d : d \mid n,\ 1\leq d<n\}$. These graphs are
highly symmetric and have important applications in chemical graph
theory. We show that $\ICG_n (D)$ has PST if and only if $n\in 4\N$
and $D=\widetilde{D_3}\cup D_2\cup 2D_2\cup 4D_2\cup \{n/2^a\}$,
where $\widetilde{D_3}=\{d\in D\ |\ n/d\in 8\N\}$, $D_2= \{d\in D\
|\ n/d\in 8\N+4\}\setminus \{n/4\}$ and $a\in\{1,2\}$. We have thus
answered the question of complete characterization of perfect state
transfer in integral circulant graphs raised in {\it Quantum
Information and Computation}, Vol. 10, No. 3\&4 (2010) 0325-–0342 by
Angeles-Canul {\it et al.} Furthermore, we also calculate perfect
quantum communication distance (distance between vertices where PST
occurs) and describe the spectra of integral circulant graphs having
PST. We conclude by giving a closed form expression calculating the
number of integral circulant graphs of a given order having PST.

\end{abstract}

\begin{description}
\item[] {\it AMS Subj. Class.}: 05C12, 05C50
\item[] {\it Keywords:} Circulant graphs; Integral graphs; Quantum spin networks; Perfect state transfer; Cayley graphs.
\end{description}

\section{Introduction}

The transfer of a quantum state from one location to another is a
crucial ingredient for many quantum information processing
protocols. There are various physical systems that can serve as
quantum channels, one of them being a quantum spin network. These
networks consist of $n$ qubits where some pairs of qubits are
coupled via XY-interaction. The perfect transfer of quantum states
from one qubit to another in such networks was first considered in
\cite{fizicarski}. There are two special qubits $A$ and $B$
representing the input and output qubit, respectively. The
transfer is implemented by setting the qubit $A$ in a prescribed
quantum state and by retrieving the state from the output qubit
$B$ after some time. The transfer is called {\it perfect state
transfer} (transfer with unit fidelity) if the initial state of
the qubit $A$ and the final state of the qubit $B$ are equal up to
a local phase rotation.

Every quantum spin network with fixed nearest-neighbor couplings
is uniquely described by an undirected graph $G$ on a vertex set
$V(G)=\{1,2,\ldots,n\}$. The edges of the graph $G$ specify which
qubits are coupled. In other words, there is an edge between
vertices $i$ and $j$ if $i$-th and $j$-th qubit are coupled.

In \cite{fizicarski} a simple XY coupling is considered  such that
the Hamiltonian of the system has the form
$$
H_G=\frac 1 2 \sum_{(i,j)\in E(G)}
\sigma_i^x\sigma_j^x+\sigma_i^y\sigma_j^y.
$$
and $\sigma_i^x,\sigma_i^y$ and $\sigma_i^z$ are Pauli matrices
acting on $i$-th qubit. The standard basis chosen for an
individual qubit is $\{|0\rangle,|1\rangle\}$ and it is assumed
that all spins initially point down ($|0\rangle$) along the
prescribed $z$ axis. In other words, the initial state of the
network is $|\underline{0}\rangle=|0_A0\ldots00_B\rangle$. This is
an eigenstate of Hamiltonian $H_G$ corresponding to zero energy.
The Hilbert space $\mathcal{H}_G$ associated with a network is
spanned by the vectors $|e_1e_2\ldots e_n\rangle$ where $e_i \in
\{0,1\}$ and, therefore, its dimension is $2^n$.

The process of transmitting a quantum state from $A$ to $B$ begins
with the creation of the initial state $\alpha
|0_A0\ldots00_B\rangle+\beta |1_A0\ldots00_B\rangle$ of the
network. Since $|\underline{0}\rangle$ is a zero-energy eigenstate
of $H_G$, the coefficient $\alpha$ will not change in time. Since
the operator of total $z$ component of the spin
$\sigma^z_{tot}=\sum_{i=1}^n \sigma_i^z$ commutes with $H_G$,
state $|1_A0\ldots00_B\rangle$ must evolve into a superposition of
the states $|i\rangle=|0\ldots01_i0,\ldots,0\rangle$ for
$i=1,\ldots,n$. Denote by $\mathcal{S}_G$ the subspace of
$\mathcal{H}_G$ spanned by the vectors $|i\rangle$,
$i=1,\ldots,n$. Hence, the initial state of network evolves in
time $t$ into the state
$$
\alpha |\underline{0}\rangle + \sum_{i=1}^n \beta_i(t) |i \rangle
\in \mathcal{S}_G.
$$
The previous equation shows that system dynamics is completely
determined by the evolution in $n$-dimensional space
$\mathcal{S}_G$. The restriction of the Hamiltonian $H_G$ to the
subspace $\mathcal{S}_G$ is an $n \times n$ matrix identical to
the adjacency matrix $A_G$ of the graph $G$.


Thus, the time evolution operator can be written in the form
$F(t)=\exp(\i A_Gt)$. The matrix exponential $\exp(M)$ is defined as
usual
$$
\exp(M)=\sum_{n=0}^{+\infty} \frac1{n!} M^n.
$$
{\it Perfect state transfer} (PST) between different vertices
(qubits) $a$ and $b$ ($1 \leq a,b \leq n$) is obtained in time
$\tau$, if $\langle a| F(t)|b \rangle=|F(\tau)_{ab}|=1$. The graph
(network) is {\it periodic} at $a$ if $|F(\tau)_{aa}|=1$ for some
$\tau$. A graph is {\it periodic} if it is periodic at each vertex
$a$.

The existence of PST for some network topologies has already been
considered in the literature. For example, Christandl et al.
\cite{fizicarski1} proved that PST occurs in paths of length one and
two between their end-vertices and also in Cartesian powers of these
graphs between vertices at maximal distance. In the recent paper
\cite{godsil}, Godsil constructed a class of distance-regular graphs
of diameter three, with PST. Some properties of quantum dynamics on
circulant graphs were studied in \cite{ahmadi03}. Saxena, Severini
and Shparlinski \cite{severini} considered circulant graphs as
potential candidates for modeling quantum spin networks having PST.
They show that a circulant graph is periodic if and only if all
eigenvalues of the graph are integers (i.e. graph is integral).
Since periodicity is a necessary condition for PST existence
\cite{severini}, circulant graphs having PST must be {\it integral
circulant graphs}. A simple and general characterization of the
existence of PST in an integral circulant graph, in terms of its
eigenvalues, was given by Ba\v si\'c, Petkovi\'c and Stevanovi\'c in
\cite{basic08a}. Furthermore, it was shown that for odd number of
vertices, there is no PST, and that among the class of unitary
Cayley graphs a subclass of integral circulant graphs, only $K_2$
(the complete graph with two nodes) and $C_4$ (the cycle of length
four) have PST. In the recent paper \cite{petkovic}, it was proven
that there exists an integral circulant graph with $n$ vertices
having PST if and only if $4 \mid n$. Several classes of integral
circulant graphs having PST were found as well and several others in
\cite{sandiego}.

In all known classes of graphs having PST {\it perfect quantum
communication distances} (i.e. the distances between vertices
where PST occurs) are considerably small compared to the order of
the graph. One idea for the distance enlargement, is to consider
networks with fixed but different couplings between qubits. These
networks correspond to graphs with weighted adjacency matrices.
For example, in \cite{fizicarski,fizicarski1} the authors showed
that PST can be achieved over arbitrarily long distances in a
weighted linear paths. Many recent papers have proposed such an
approach \cite{iranci,sandiego,sandiego1}.


Studying PST in integral circulant graphs can also be interpreted
as a contribution to the spectral theory of integral graphs. These
graphs are highly symmetric and have some remarkable properties
connecting graph theory and number theory. The term 'integral
circulant graph' first appears in the work of So \cite{wasin},
where a nice characterization of these graphs in terms of their
symbol set is given. The upper bounds on the number of vertices
and the diameter of integral circulant graphs were given in
\cite{severini}. Furthermore, Stevanovi\'c, Petkovi\'c and Ba\v
si\'c \cite{stevanovic08} improved the upper bound. Various other
properties of unitary Cayley graphs were recently investigated.
For example, Berrizbeitia and Giudici \cite{berrizbeitia04} and
Fuchs \cite{fuchs05} established the lower and upper bound on the
size of the longest induced cycle. Klotz and Sander \cite{klotz}
determined the diameter, clique number, chromatic number and
eigenvalues of unitary Cayley graphs. Ba\v si\'c and Ili\' c
\cite{basic08} calculated the clique number of integral circulant
graphs with exactly one and two divisors and also provided an
inequality for the general case.

In this paper we proceed with the study of circulant networks
supporting PST initiated in \cite{basic2, basic08a, sandiego,
petkovic, severini}.
First we give some properties of the spectra of
$\ICG_n(\widetilde{D_1})$ where $\widetilde{D_1}=\{d\in D\ |\ 4\nmid
n/d\}$. In that subsection we present some preliminary results which
are used in the sequel of the section. We show that $\ICG_n (D)$ has
PST if and only if $n\in 4\N$ and $D=\widetilde{D_3}\cup D_2\cup
2D_2\cup 4D_2\cup \{n/2^a\}$, where $\widetilde{D_3}= \{d\in D\ |\
n/d\in 8\N\}$, $D_2= \{d\in D\ |\ n/d\in 8\N+4\}\setminus \{n/4\}$
and $a\in\{1,2\}$. Using the previous result we prove that perfect
quantum communication distance
is either equal to one if $n/2\in D$ or equal to two if $n/4\in D$.
Moreover, we describe the spectra of integral circulant graphs
having PST. The paper is concluded with a formula for the number of
$\ICG_n(D)$ having PST as a function of the number of vertices $n$.
This results answer questions posed in
\cite{sandiego,petkovic,severini}.


\section{Integral circulant graphs}

The {\it circulant graph} $G(n;S)$ is a graph on vertices
$\Z_n=\{0,1,\ldots,n-1\}$ such that each vertex $i$ is adjacent to
vertices $i+_n s$ for all $s\in S$. The set $S \subseteq \Z_n$ is
called the {\it symbol} of the graph $G(n;S)$ and $+_n$ denotes
addition modulo $n$. Note that the degree of $G(n;S)$ is $\#S$. A
graph is {\it integral} if all its eigenvalues are integers. Wasin
So has characterized integral circulant graphs \cite{wasin} in the
following theorem:

\begin{thm}
\label{th:wasin} {\bf \cite{wasin}} A circulant graph $G(n;S)$ is
integral if and only if
$$
S=\bigcup_{d \in D} G_n(d),
$$
for some set of divisors $D \subseteq D_n$. Here $G_n(d)=\{ k \ : \
\gcd(k,n)=d, \ 1\leq k \leq n-1 \}$, and $D_n$ is the set of all
divisors of $n$, different from $n$.
\end{thm}

Therefore an {\it integral circulant graph} (ICG) $G(n;S)$ is
defined by its order $n$ and the set of divisors $D$. Such graphs
are also known as {\it gcd-graphs} (see for example \cite{klotz}).
An integral circulant graph with $n$ vertices, defined by the set of
divisors $D \subseteq D_n$ will be denoted by $\ICG_n(D)$. From
Theorem \ref{th:wasin} we have that the degree of an integral
circulant graph is $\deg \ICG_n(D)=\sum_{d \in D}\varphi(n/d). $
Here $\varphi(n)$ denotes the Euler-phi function \cite{HardyWright}.

The eigenvalues and eigenvectors of $\ICG_n(D)$ are given in
\cite{severini} as \bb \lambda_j=\sum_{s \in S} \omega^{js}_n,
\quad v_j=[1 \ \omega_n^s \ \omega_n^{2s} \cdots
\omega_n^{(n-1)s}], \label{eq:eigens} \ee where
$\omega_n=\exp(\i2\pi/n)$ is the $n$-th root of unity. Denote by
$c(n,j)$ the following expression
\bb
c(j,n)=\mu(t_{n,j})\frac{\varphi(n)}{\varphi(t_{n,j})}, \quad
t_{n,j}=\frac n{\gcd(n,j)},
 \label{ramanujan}
\ee where $\mu$ is the M\" obius function defined as

\begin{eqnarray}
\mu(n)&=&\left\{
\begin{array}{rl}
1, &  \mbox{if}\ n=1  \\
0, & \mbox{if $n$ is not square--free} \\
(-1)^k, & \mbox {if $n$ is product of $k$ distinct prime numbers}.
\end{array} \right.
\end{eqnarray}
The expression $c(j,n)$ is known as the {\it Ramanujan function}
(\cite[p.~55]{HardyWright}). Eigenvalues $\lambda_j$ can be
expressed in terms of the Ramanujan function as follows
(\cite{klotz}, Theorem 16) \bb \lambda_j=\sum_{d\in D} c(j,n/d).
\label{ldef} \ee Let us observe that the Ramanujan function has the
following basic properties which we will make use of in the paper.

\begin{prop} \label{prop:c} For any positive integers $n,j$ and $d$ such that $d \mid n$, holds
\begin{eqnarray}
c(0,n)&=&\varphi(n), \\
c(1,n)&=&\mu(n),\\
c(2,n)&=&\left\{
\begin{array}{rl}
\mu(n), &  n \in 2\N+1  \\
\mu(n/2), & n \in 4\N+2 \\
2\mu(n/2), & n \in 4\N
\end{array} \right.
\\
c(n/2,n/d)&=&\left\{ \begin{array}{rl}
\varphi(n/d), & d \in 2\N \\
-\varphi(n/d), & d \in 2\N+1 \\
\end{array}\right.
\end{eqnarray}
\end{prop}
\begin{dok}
Directly using relation (\ref{ramanujan}).
\end{dok}

The integral circulant graph $\ICG_n(D)$ is connected if and only
if $\gcd(n,d_1,\ldots,d_k)=1$ where $D=\{d_1,\ldots,d_k\}$. In the
rest of the paper we will only consider connected integral
circulant graphs.

\section{Perfect state transfer}

Let $G$ be an undirected graph and denote by $A_G$ its adjacency
matrix. Let $F(t)=\exp(iA_Gt)$. There is a {\it perfect state
transfer} (PST) in graph $G$ \cite{fizicarski,godsil,severini} if
there are distinct vertices $a$ and $b$ and a positive real number
$t$ such that $|F(t)_{ab}|=1$.

Let $\lambda_0,\lambda_2,\ldots,\lambda_{n-1}$ be the eigenvalues (not
necessarily distinct) of $A_G$ and $u_0,u_1,\ldots,u_{n-1}$ be the
corresponding normalized eigenvectors. We use spectral decomposition
of the real symmetric matrix $A_G$ (see for example
\cite{godsilknjiga} (Theorem 5.5.1) for more details). The matrix
function $F(t)$ can be represented as

\bb F(t)=\sum_{k=0}^{n-1}
\exp(\i \lambda_k t) u_k u_k^*. \label{eq:H} \ee

Now let
$G=\ICG_n(D)$ be an integral circulant graph. By simple calculation
and using (\ref{eq:eigens}), we see that $\|v_k\|=\sqrt{n}$ and
hence $u_k=v_k / \sqrt{n}$. Expression (\ref{eq:H}) now becomes
$$
F(t)=\frac1n \sum_{k=0}^{n-1} \exp(\i \lambda_k t) v_k v_k^*.
$$
In particular, from the last expression and (\ref{eq:eigens}) it
directly follows
$$
F(t)_{ab}=\frac1n \sum_{k=0}^{n-1} \exp(\i \lambda_k t)
\omega_n^{k(a-b)}.
$$
This expression is given in \cite{severini} (Proposition 1).
Finally, our goal is to check whether there exist distinct integers
$a,b \in \Z_n$ and a positive real number $t$ such that
$|F(t)_{ab}|=1$, i.e. \bb \left|\frac1n\sum_{k=0}^{n-1} \exp(\i
\lambda_k t) \omega_n^{k(a-b)}\right|=1. \label{PSTdef} \ee Since
the left-hand side of (\ref{PSTdef}) depends on $a$ and $b$ only as
a function of $a-b$ we can, without any loss of generality, assume
that $b=0$. Therefore, throughout the paper we consider the
existence of PST only between vertices $a$ and $0$.

We restate some results proved in \cite{basic08a}. These results
establish necessary and sufficient conditions for (\ref{PSTdef}).

\begin{thm} {\bf \cite{basic08a}}  \label{th:PST1}
There exists PST in $\ICG_n(D)$ between vertices $a$ and $0$
if and only if there are integers $p$ and $q$ such that
$\gcd(p,q)=1$ and \bb \frac{p}q(\lambda_{j+1}-\lambda_j)+\frac{a}n
\in \Z, \label{PSTr2} \ee for all $j=0,\ldots,n-2$.
\end{thm}



\begin{thm} {\bf \cite{basic08a}}
\label{th:odd} There is no PST in $\ICG_n(D)$ if $n/d$ is odd for
every $d \in D$. For $n$ even, if there exists PST in
$\ICG_n(D)$ between vertices $a$ and $0$, then $a=n/2$.
\end{thm}

According to Theorem \ref{th:odd}, PST may exist in $\ICG_n(D)$
only between vertices $n/2$ and $0$ (i.e., between $b$ and $n/2+b$
as mentioned in \cite{severini}). Hence we will avoid referring to
the input and output vertex and will just say that there exists
PST in $\ICG_n(D)$.

The next corollary is derived from Theorem \ref{th:PST1} and is
further used as the criterion for a nonexistence of PST.

\begin{cor} {\bf \cite{basic08a}}
\label{cor:eql} If $\lambda_j=\lambda_{j+1}$ for some
$j=0,\ldots,n-2$ then there is no PST in $\ICG_n(D)$.
\end{cor}



For a given prime number $p$ and an integer $n\in
\N_0$\footnote{$\N_0=\N\cup\{0\}$}, denote by $S_p(n)$ the maximal
number $\alpha$ such that $p^{\alpha} \mid n$ if $n\in \N$, and
$S_p(0)=+\infty$. The following result was proven in \cite{basic08a}
and is further used as a criterion for the existence of PST.

\begin{lem} {\bf \cite{basic08a}} \label{lem:pow2}
There exists PST in $\ICG_n(D)$, if and only if there exists a
number $m \in \N_0$ such that the following holds for all
$j=0,1,\ldots, n-2$ \bb S_2(\lambda_{j+1}-\lambda_j)=m.
\label{PSTr4} \ee
\end{lem}

The following corollary follows directly from Lemma \ref{lem:pow2}.

\begin{cor} \label{cor:par}
Let $\ICG_n(D)$ have PST. One of the following two statements must
hold
\begin{itemize}
\item[{\bf 1.}] $\lambda_j \equiv \lambda_{j+1} \pmod 2$ for every
$0 \leq j \leq n-1$ (i.e., all eigenvalues $\lambda_j$ have the same
parity).
\item[{\bf 2.}] $\lambda_j \equiv \lambda_{j+1}+1 \pmod 2$ for every
$0 \leq j \leq n-1$ (i.e., $\lambda_j$ are alternatively odd and
even).
\end{itemize}
\end{cor}




We end this section with the following result concerning
nonexistence of PST in $\ICG_n(D)$, where $n\in 4\N+2$.
\begin{thm} {\bf \cite{petkovic}} There is no PST in $\ICG_n(D)$ for an arbitrary set of divisors $D$ for $n\in 4\N+2$.
\label{thm:4np2}
\end{thm}

\section{Integral circulant graphs having PST}




Let $ICG_n(D)$ be an arbitrary integral circulant graph. We define
sets $D_i\subseteq D$ for $0\leq i\leq l$, where $l=S_2(n)$, in the
following way

$$
D_i=\{d\in D\ |\ S_2(n/d)=i\}.
$$

For simplicity of notation we also define sets
$\widetilde{D_1},\widetilde{D_3}\subset D$ to be
$\widetilde{D_1}=D_0\cup D_1$ and $\widetilde{D_3}=\cup_{i\geq 3} \
D_i$.

Let us introduce the notation $kD$ for the set $\{kd\ |\ d\in D\}$
for a positive integer $k$.

\subsection{Spectrum of the integral circulant graph $\ICG_n(\widetilde{D_1})$}

In this section we deal with integral circulant graphs $\ICG_n(D)$
such that for each divisor $d\in D$ it holds that $4\nmid \frac n
d$. In the rest of the section we will consider only such classes of
graphs unless otherwise stated.

In Lemmas \ref{lem:nep} and \ref{lem:cpm} we present some properties of the Ramanujan function.

Throughout the section, we let $n = 2^{\alpha_0}p_1^{\alpha_1}
p_2^{\alpha_2} \cdot \ldots \cdot p_k^{\alpha_k}$, where $p_1 < p_2
< \ldots < p_k$ are distinct primes, and $\alpha_i \geqslant 1$ for
$1\leq i\leq k$ and $\alpha_0\geq 0$.

\begin{lem}
For $n\geq 2$ it holds that  $c(j,n) \in 2\N+1$ if and only if $4
\nmid n$ and $j=p_1^{\alpha_1-1} \cdots p_k^{\alpha_k-1}J$ for some
integer $J$ such that $\gcd(J,n) \in \{1,2\}$. \label{lem:nep}
\end{lem}

\begin{dok}

\bigskip

\noindent $(\Rightarrow :)$ Suppose that $c(j,n)$ is an odd integer.
Since $c(j,n)=\mu(t_{n,j}) \varphi(n)/\varphi(t_{n,j})$, it holds
that $\mu(t_{n,j})=\pm 1$, i.e. $t_{n,j}$ is square-free and
$\varphi(n)/\varphi(t_{n,j})$ is an odd integer.

Suppose that for some odd $p_i$ it holds that $p_i \nmid t_{n,j}$.
Let $n'={n}/{p_i^{\alpha_i}}$. Since $t_{n,j} \mid n'$ and so
$\varphi(t_{n,j}) \mid \varphi(n')$ we obtain that
$$
c(j,n)=\pm\frac{\varphi(n)}{\varphi(t_{n,j})}=\pm\frac{\varphi(p_i^{\alpha_i})\varphi(n')}{\varphi(t_{n,j})}=\pm
p_i^{\alpha_i-1}(p_i-1)\frac{\varphi(n')}{\varphi(t_{n,j})}.
$$
The last equation implies that $c(j,n)$ is even since $p_i-1$ is
even. This is a contradiction and we can conclude that $p_i \mid
t_{n,j}$ for every $1\leq i\leq k$.

Now we have that $\varphi(t_{n,j})=(p_1-1)\cdots (p_k-1)$ and thus
$$
c(j,n)=2^{\alpha_0-1}p_1^{\alpha_1-1}p_2^{\alpha_2-1}\cdots
p_k^{\alpha_k-1}.
$$
Since $c(j,n)$ is odd it holds that $0\leq \alpha_0\leq 1$ or
equivalently $4\nmid n$.

If $n \in 2\N+1$ it must hold that $t_{n,j}=p_1 \cdots p_k$ since
$t_{n,j}$ is square-free. If $n \in 4\N+2$, we have two
possibilities for $t_{n,j}$, $t_{n,j}=p_1 \cdots p_k$ or
$t_{n,j}=2p_1 \cdots p_k$ depending on the parity of $j$. 

Furthermore, using $n=\gcd(n,j) t_{n,j}$ we obtain that
$\gcd(n,j)=p_1^{\alpha_1-1} \cdots p_k^{\alpha_k-1}$ ($t_{n,j}$ and
$n$ have the same parity) or $\gcd(n,j)=2p_1^{\alpha_1-1} \cdots
p_k^{\alpha_k-1}$ (otherwise). This implication of the lemma is now
straightforward.

\smallskip

\noindent $(\Leftarrow :)$ Since $\gcd(n,j)=p_1^{\alpha_1-1} \cdots
p_k^{\alpha_k-1}\gcd(J,n)$ and $\gcd(J,n)=\{1,2\}$, it holds that
$t_{n,j}=p_1 \cdots p_k$ or $t_{n,j}=2p_1 \cdots p_k$. In either
case it holds that $\varphi(t_{n,j})=(p_1-1)\cdots (p_k-1)$. Now
since
$$
c(j,n)=\mu(t_{n,j})\frac{\varphi(n)}{\varphi(t_{n,j})}=\pm
p_1^{\alpha_1-1} \cdots p_k^{\alpha_k-1},
$$
we conclude that $c(j,n) \in 2\N+1$.
\end{dok}

\begin{lem} Let $d$ be an arbitrary divisor of $n$ such that $n/d\in 2\N+1$ and $0\leq j \leq n-1$ be an arbitrary integer, then $c(j,n/d)=-c(j,2n/d)$ for $j\in 2\N+1$ and $c(j,n/d)=c(j,2n/d)$ for $j\in 2\N$.
\label{lem:cpm}
\end{lem}
\begin{dok}
As $n/d\in 2\N+1$ we conclude that $\varphi(2n/d)=\varphi(n/d)$.

Suppose that $j \in 2\N+1$. Then $\gcd(2n/d,j)=\gcd(n/d,j)$ and
$$
t_{2n/d,j}=\frac{2n}{d\gcd(2n/d,j)}=2\frac{n}{d\gcd(n/d,j)}=2t_{n/d,j}.
$$
Furthermore, it holds that $\varphi(t_{2n/d,j})=\varphi(t_{n/d,j})$
since $t_{n/d,j}$ is odd. Also we have that $t_{2n/d,j}$ is
square-free if and only if $t_{n/d,j}$ is square-free, and
$\mu(t_{2n/d,j})=-\mu(t_{n/d,j})$. Now we can directly conclude that
$$
c(j,2n/d)=\mu(t_{2n/d,j})\frac{\varphi(2n/d)}{\varphi(t_{2n/d,j})}=-\mu(t_{n/d,j})\frac{\varphi(n/d)}{\varphi(t_{n/d,j})}=-c(j,n/d).
$$

Now suppose that $j \in 2\N$. We have $\gcd(2n/d,j)=2\gcd(n/d,j)$
and also $t_{2n/d,j}=t_{n/d,j}$. This yields directly  that
$c(j,2n/d)=c(j,n/d)$.
\end{dok}

\begin{thm}
\label{tmh:oneodd}For an arbitrary integral circulant graph
$ICG_n(D)$ there exists an odd number $0 \leq j \leq n-1$ with
$\lambda_j$ also odd if and only if there is a divisor $d\in D$
satisfying $d/2 \notin D$ and $2d \notin D$. \label{lem:oneodd}
\end{thm}
\begin{dok}

\medskip

\noindent $(\Rightarrow:)$ Suppose that for every $d \in D$ it holds
that either $2d\in D$ or $d/2 \in D$. It follows that $D=D_1\cup
2D_1$. Let $j \in 2\N+1$. According to Lemma \ref{lem:cpm} we have
$c(j,n/d)=-c(j,2n/d)$ for any $d \in D$ such that $n/d \in 2\N+1$
and therefore
$$
\lambda_j=\sum_{d\in D_1\cup 2D_1} c(j,n/d)=\sum_{d\in D_1}
c(j,n/d)+ c(j,2n/d)=0\in 2\N.
$$ 
We conclude that all eigenvalues with odd indices are even.
\medskip

\noindent $(\Leftarrow:)$ Let $D''_1=\{d\in D\ |\
n/d\in4\N+2\Rightarrow\ 2d\in D\}$ and $D''=D''_1\cup 2D''_1$.


Let $D'=D \setminus D''$. By the assumption, $D'$ is a non-empty
set. According to Lemma \ref{lem:cpm} it holds that
$c(j,n/d)+c(j,2n/d) \in 2\N$ for every $d \in D''$ such that $n/d\in
2\N+1$. Denote by $d'_{max}=\max D'$. Since $d'_{max}$ is a divisor
of $n$, it can be represented in the form
$d'_{max}=2^{\beta_0}p_1^{\beta_1} \cdots p_k^{\beta_k}$ where $0
\leq \beta_i \leq \alpha_i$ for $i=1,\ldots,k$. Without loss of
generality, we can suppose that there exists $1 \leq s \leq k$ such
that $\beta_i<\alpha_i$ for $i=1,\ldots,s$ and $\beta_i=\alpha_i$
for $i=s+1,\ldots,k$. Then we can write
$n/d'_{max}=2^{\alpha_0-\beta_0} p_1^{\alpha_1-\beta_1} \cdots
p_s^{\alpha_s-\beta_s}$. Denote by
$$
j_0=p_1^{\alpha_1-\beta_1-1} \cdots p_s^{\alpha_s-\beta_s-1}
p_{s+1}^{\alpha_{s+1}} \cdots p_k^{\alpha_k}.
$$
It holds trivially that $0 \leq j_0 \leq n-1$. Lemma \ref{lem:nep}
yields directly that $c(j_0,n/d'_{max})$ is odd since $n/d'_{max}
\geq 2$.

Suppose that $D' \setminus \{d'_{max}\} \neq \emptyset$ and let $d
\in D' \setminus \{d'_{max}\}$ be an arbitrary divisor with its
prime factorization $d=2^{\gamma_0} p_1^{\gamma_1} \cdots
p_k^{\gamma_k}$.

We will show that there exists $1 \leq i \leq k$ such that $0 \leq
\gamma_i < \beta_i \leq \alpha_i$.  Suppose this is not the case,
which means that $0 \leq \beta_i\leq \gamma_i  \leq \alpha_i$ for
$1\leq i\leq k$. If $ \beta_0\leq\gamma_0$ then $d'_{max} \mid d,$
which is a contradiction. Similarly, if $ \beta_0=\gamma_0+1$ then
it holds that $d'_{max} \mid 2d$, which implies $d=d'_{max}/2$,
providing a contradiction with the definition of the set $D'$.

Let $i$ be an arbitrary index such that $\gamma_i<\beta_i$. Suppose that $\alpha_i-\beta_i \geq 1$. Then $i\leq s$ and
$S_{p_i}(n/d)=\alpha_i-\gamma_i \geq 2$. Since
$S_{p_i}(j_0)=\alpha_i-\beta_i-1>\alpha_i-\gamma_i-1=S_{p_i}(n/d)-1$, from Lemma
\ref{lem:nep} we can conclude that $c(j_0,n/d)$ is even.

Now suppose that $\alpha_i=\beta_i$. Then $i>s$ and
$S_{p_i}(n/d)=\alpha_i-\gamma_i \geq 1$. Again since
$S_{p_i}(j_0)=\alpha_i>\alpha_i-\gamma_i-1=S_{p_i}(n/d)-1$, Lemma \ref{lem:nep}
yields that $c(j_0,n/d)$ is even.

\smallskip

This implies that there is an odd index $j_0$ such that
$c(j_0,n/d'_{max})$ is odd and $c(j_0,n/d)$ is even for every $d \in
D' \setminus \{d'_{max}\}$. Now we have
$$
\lambda_{j_0}=c(j_0,n/d'_{max})+ \sum_{d\in D' \setminus
\{d'_{max}\}} c(j_0,n/d)+ \sum_{d \in D''}c(j_0,n/d) \in 2\N+1,
$$
since both sums in the last expression are even.

If $d_{max}$ is the only divisor contained in $D$, the above sum is
reduced to
$$
\lambda_{j_0}=c(j_0,n/d'_{max})+  \sum_{d \in D''}c(j_0,n/d) \in
2\N+1,
$$
and it is still even.
\end{dok}

Let us mention two direct consequences of the previous theorem.
The first one is actually the contrapositive of the assertion of
the theorem.

\begin{lem}
\label{lem:alleven} All eigenvalues of $\ICG_n(D)$ on odd positions
are even  if and only if $D=D_1\cup 2D_1$.
\end{lem}

\begin{lem}
\label{lem:allodd} Let $n/2\in D$. All eigenvalues of $\ICG_n(D)$ on
odd positions are odd if and only if $D=D^*_1\cup 2D^*_1\cup
\{n/2\}$ where $D^*_1=D_1\setminus\{n/2\}$.
\end{lem}
\begin{dok}
Suppose that all eigenvalues of $\ICG_n(D)$ on odd positions are
odd. Denote by $\lambda_j$ the eigenvalues of the graph. Now,
consider the integral circulant graph $\ICG_n(D')$ where $D'=D
\setminus \{n/2\}$. Denote by $\lambda'_j$ the eigenvalues of
integral circulant graph $\ICG_n(D')$. Since \bb
t_{2,j}=\frac2{\gcd(2,j)}= \left\{
\begin{array}{rl}
2, & 2 \nmid j\\
1, & 2 \mid j
\end{array} \right., \quad
c(j,2)=\left\{ \begin{array}{rl}
-1, & 2 \nmid j\\
1, & 2 \mid j
\end{array} \right..
\label{cj2} \ee it holds that

\bb \label{cj2final} \lambda_j=\left\{
\begin{array}{rl}
\lambda'_j+1,& 2 \mid j \\
\lambda'_j-1,& 2 \nmid j
\end{array} \right..
\ee

This gives that all eigenvalues of $\ICG(D')$ on odd positions are
even. According to Lemma \ref{lem:alleven} we have that
$D'=D^*_1\cup 2D^*_1$ where $D^*_1=D_1\setminus\{n/2\}$ which
completes the first part of the proof.

The converse of the assertion can be proven analogously. Let
$D=D^*_1\cup 2D^*_1\cup \{n/2\}$ and $D'=D\setminus\{n/2\}$.
According to Lemma \ref{lem:alleven} all eigenvalues of $\ICG_n(D')$
on odd positions are even. Furthermore, using the relation
(\ref{cj2final}) we have that all eigenvalues of $\ICG_n(D)$ on odd
positions are odd.
\end{dok}

\subsection{Perfect quantum distance, spectrum and characterization  of $\ICG_n(D)$ having PST}

The main result of this section is characterization of $\ICG_n(D)$
having PST. In addition, we also calculate perfect quantum distance
of integral circulants having PST. Thus, according to Theorem
\ref{thm:4np2} from now on we assume that $n$ is divisible by four.

\begin{thm} \label{thm:par}
Let $\ICG_n(D)$ have PST. 
If $n/2 \in D$ then all eigenvalues are odd, otherwise they are
even. Moreover, in the first case the eigenvalues on odd positions
are equal to $-1$ and in  the second one the eigenvalues on odd
positions are equal to $0$.
\end{thm}
\begin{dok}
According to Proposition \ref{prop:c} we have $\lambda_1=\sum_{d \in
D} \mu(n/d)$. Since $4\mid n/d$ for $d\in D_2\cup \widetilde{D_3}$
we conclude that $\mu(n/d)=0$ and therefore $\lambda_1=\sum_{d \in
\widetilde{D_1}} \mu(n/d)$. Using Proposition \ref{prop:c} once
again we see that $\lambda_2=\sum_{d \in D_0} \mu(n/d)+\sum_{d \in
D_1} \mu(n/2d)+\sum_{d \in D_2\cup \widetilde{D_3}} 2\mu(n/2d)$. For
$d\in \widetilde{D_3}$ we have $4\mid n/2d$, which yields
$$
\lambda_2-\lambda_1=\sum_{d \in D_1} (\mu(n/2d)-\mu(n/d))+ 2\sum_{d
\in D_2}\mu(n/2d)\in 2\N.
$$

By Lemma \ref{lem:pow2} all the differences
$\lambda_{i+1}-\lambda_i\in 2\N$ for $0\leq i\leq n-2$, since
$\ICG_n(D)$ has PST. If $n/2\notin D$ then $\lambda_0\in 2N$ and
thus all the eigenvalues are even, otherwise $\lambda_0\in 2N+1$
and all the eigenvalues are odd.

Let $j\in 2\N+1$. For $d\in D_2\cup \widetilde{D_3}$ we have that
$4\mid t_{n/d,j}$ and thus $c(j,n/d)=0$. This reduces the formula
for the $j$-th eigenvalue to
$$\lambda_j=\sum_{d \in \widetilde{D_1}} c(j,n/d).$$ The condition $n/2\notin D$
yields that $\lambda_0\in 2\N$ and according to the first part of
the proof all the eigenvalues are even.

Let $\mu_j$, $0\leq j\leq n-1$,  be the eigenvalues of the integral
circulant graph $\ICG_n(\widetilde{D_1})$. Thus $\lambda_j=\mu_j$
for any odd $0\leq j\leq n-1$, so $\mu_j\in 2\N$ for $j\in 2\N+1$.
From Lemma \ref{lem:alleven} we conclude that all the eigenvalues
$\mu_j$ on odd positions are even if and only if $D_0=2D_1$ and
$\widetilde{D_1}=D_1\cup 2D_1$. By the first part of the proof of
Theorem \ref{lem:oneodd} we deduce that $\mu_j=0$ for $j\in 2\N+1$
and consequently  $\lambda_j=0$ for $j\in 2\N+1$.

Analogously, if $n/2\in D$ we obtain that $\lambda_j\in 2\N+1$ for
$0\leq j\leq n-1$ and so $\mu_j\in 2\N+1$ for odd $0\leq j\leq n-1$.
Now, Lemma \ref{lem:allodd} yields $D=D^*_1\cup 2D^*_1\cup \{n/2\}$
where $D^*_1=D_1\setminus\{n/2\}$.

Now, consider the integral circulant graph $\ICG_n(D')$ where
$D'=D \setminus \{n/2\}=D^*_1\cup 2D^*_1$. Denote by $\lambda'_j$
the eigenvalues of the integral circulant graph $\ICG_n(D')$. From
(\ref{cj2final}) we obtain that $\lambda_j=\lambda'_j-1$ for $j\in
2\N+1$. But, according to the first part of the proof we have that
$\lambda'_j=0$ for $j\in 2\N+1$ and so $\lambda_j=-1$ for $j\in
2\N+1$.

\end{dok}

From the proof of the last theorem we can derive the following
important corollary

\begin{cor}
\label{cor:1} If $\ICG_n(D)$ has PST then  $D_0=2(D_1\setminus
\{n/2\})$.
\end{cor}

Using Theorem \ref{thm:par} we can establish a more precise
criterion for the characterization of integral circulant graphs
having PST, than the one given by Corollary \ref{cor:par}.

\begin{lem}
\label{criterion1}  $\ICG_n(D)$ has PST if and only if there
exists an integer $k\geq 1$ such that one of the following
conditions are satisfied
\begin{itemize}
\item[{\bf i)}] $S_2(\lambda_{2j})=k$ and $\lambda_{2j+1}=0$, if $n/2 \not \in D$
\item[{\bf ii)}] $S_2(\lambda_{2j}+1)=k$ and $\lambda_{2j+1}=-1$, if $n/2  \in D$
\end{itemize}
for $0\leq j\leq  n/2$.
\end{lem}

\begin{lem}
\label{lem:c(j,n/d)} Let $n$ be an even number, $d$ be a divisor of
$n$ and $n_1=n/2$. For an even number $0\leq j\leq n-1$ the
following equalities hold
\begin{itemize}
\item[{\bf 1.}] $c(j,n/d)=c(j/2,\frac {n_1} {d/2})$, if $d\in D_0$
\item[{\bf 2.}] $c(j,n/d)=c(j/2,n_1 / d)$, if $d\in D_1$
\item[{\bf 3.}] $c(j,n/d)=2c(j/2,n_1 / d)$, if $d\in D_2\cup \widetilde{D_3}$.
\end{itemize}
\end{lem}

\begin{dok}
\begin{itemize}
\item[{\bf 1.}] Suppose that $d \in D_0$. Then
$\gcd(n/d,j)=\gcd(\frac {n_1} {d/2},j/2)$ and
$$
t_{n/d,j}=\frac{n}{d\gcd(n/d,j)}=\frac{2n_1}{2\frac d
2\gcd(\frac{n_1}{d/2} ,j/2)}=t_{\frac {n_1} {d/2},j/2}.
$$
Furthermore, it holds that $\varphi(n/d)=\varphi(\frac {n_1}{d/2})$
and so $c(j, n/d)=c(j/2,\frac {n_1} {d/2})$.

\item[{\bf 2.}] Suppose now $d \in D_1$. Then
$\gcd(n/d,j)=2\gcd(n_1/d,j/2)$ and
$$
t_{n/d,j}=\frac{n}{d\gcd(n/d,j)}=\frac{2n_1}{d\ 2\gcd(n_1/d
,j/2)}=t_{n_1/d,j/2}.
$$
We also conclude that $\varphi(n/d)=\varphi(2\frac
{n_1}{d})=\varphi(n_1/d)$ and so $c(n/d, j)=c(n_1 / d, j/2)$.

\item[{\bf 3.}] Suppose now $d \in D_2\cup \widetilde{D_3}$. By the preceding
case we obtain that $\gcd(n/d,j)=2\gcd(n_1/d,j/2)$ and $
t_{n/d,j}=t_{n_1/d,j/2}.$ Let $k=S_2(n/d)$ and $n'$ be an integer
such $n/d=2^k\ n'/d$. Notice that $k\geq 2$, since $d\in D_2\cup
\widetilde{D_3}$. Also, $n'/d$ is odd and thus $\gcd(2^k,n'/d)=1$
which yields $\varphi(n/d)=\varphi(2^k\frac
{n'}{d})=2^{k-1}\varphi(\frac {n'}{d})$.

Furthermore we have
$$\varphi(n/d)=
2^{k-1}\varphi(\frac {n'}{d})=2\varphi(2^{k-1})\varphi(\frac
{n'}{d})=2\varphi(2^{k-1}\frac {n'}{d})=2\varphi(n_1/d)$$ and so
$c(j,n/d)=2c(j/2,n_1/d)$, if $d\in D_2\cup \widetilde{D_3}$.
\end{itemize}
\end{dok}







\begin{lem}
\label{lem:D''1=2D_2} If $\ICG_n(D)$ has PST then
$D_1=2(D_2\setminus \{n/4\})$.
\end{lem}
\begin{dok}

 Let $\lambda_j$ be an eigenvalue of $\ICG_n(D)$ where $0\leq
j\leq n-1$.

According to Corollary \ref {cor:1} it holds that
$D_0=2(D_1\setminus \{n/2\})$ and so
$$\lambda_j=\sum_{d\in 2(D_1\setminus
\{n/2\})}c(j,n/d)+\sum_{d\in D_1}c(j,n/d)+\sum_{d\in D_2\cup
\widetilde{D_3}}c(j,n/d).
$$ From the relation
(\ref{cj2}) we have $c(j,2)=1$ for $j\in 2\N$.

For $d\in 2(D_1\setminus\{n/2\})$, let $d=2d'$ where $d'\in
D_1\setminus\{n/2\}$. Then $c(j, n/d)=c(j,n_1/d')$ and according to
Lemma \ref{lem:c(j,n/d)} (part 1.), we have
$c(j,n_1/d')=c(j/2,\frac{n_1/2}{d'/2})=c(j/2, n1/d')$.

If $d\in D_1$ using Lemma \ref{lem:c(j,n/d)} (part 2.) it holds that
$c(j,n/d)=c(j/2, n1/d)$.

Finally, if $d\in D_2\cup \widetilde{D_3}$ using Lemma
\ref{lem:c(j,n/d)} (part 3.) it holds that $c(j,n/d)=2c(j/2, n1/d)$.

Taking the discussion above  into account we obtain

\bb \lambda_j= \left\{ \begin{array}{rl}
c(j,2)+2\sum_{d\in D_1\setminus \{n/2\}\cup D_2\cup \widetilde{D_3}}c(j/2,n_1/d)=2\lambda'_{j/2}+1, & n/2\in D\\
2\sum_{d\in D_1\setminus \{n/2\}\cup D_2\cup
\widetilde{D_3}}c(j/2,n_1/d)=2\lambda'_{j/2}, & n/2\not \in D
\end{array} \right.
\ee where $n_1=n/2$ and $\lambda'_i$ are the eigenvalues of the
integral circulant graph $\ICG_{n_1}(D')$ where $D'=(D_1\setminus
\{n/2\})\cup D_2\cup \widetilde{D_3}$.

We conclude that $\gcd(n/d,j)=2$ for $j\in 4\N+2$ and $d\in
\widetilde{D_3}$, hence that $4\mid t_{n/d,j}$, and finally that
$c(j,n/d)=0$. This yields that \bb\lambda_j=\sum_{d\in
\widetilde{D_1}}c(j,n/d)+\sum_{d\in D_2}c(j,n/d)=\left\{
\begin{array}{rl}
2\lambda'_{j/2}+1, & n/2\in D\\
2\lambda'_{j/2}, & n/2\not\in D
\end{array} \right.
\ee where $\lambda'_j=\sum_{d\in
D_1\setminus\{n/2\}}c(j,n_1/d)+\sum_{d\in D_2}c(j,n_1/d)$. This
means that the eigenvalues $\lambda'_{j}$ for odd $0\leq j\leq
n_1-1$ coincide with the eigenvalues on odd positions of
$\ICG_{n_1}(D_1\cup D_2\setminus\{n/2\})$. Let us denote by $\mu_j$
the eigenvalues of $\ICG_{n_1}(D_1\cup D_2\setminus \{n/2\})$, for
$0\leq j\leq n_1-1$.

\medskip

From Lemma \ref{criterion1} it follows that $S_2(\lambda_j)=k$ or
$S_2(\lambda_j+1)=k$, for $j\in 4\N+2$ and some integer $k\geq 1$,
depending on whether $n/2\not\in D$ or $n/2\in D$. Furthermore, we
have that either $S_2(\lambda'_{j/2})=k-1$ or
$S_2(\lambda'_{j/2}+1)=k-1$ for odd $0\leq j/2\leq n_1-1$. Since
$\mu_j=\lambda'_j$ for odd $0\leq j\leq n_1-1$ (eigenvalues $\mu_j$
on odd positions have the same parity), Lemma \ref{lem:alleven} and
Lemma \ref{lem:allodd} yield that $D_1=2(D_2\setminus\{n_1/2\})$,
which completes the proof.








\end{dok}

\begin{thm}
\label{thm:n/2n/4} If $\ICG_n(D)$ has PST then either $n/2\in D$ or
$n/4\in D$.
\end{thm}
\begin{dok}

\noindent {\bf Case 1.}  $n/2\not \in D.$ Suppose also that $n/4\not
\in D.$ Since $n/2 \notin D$, according to Corollary \ref {cor:1} it
holds that $D_0=2D_1$ and using Proposition \ref{prop:c} we have $
\lambda_2=\sum_{d\in 2D_1}\mu(n/d)+\sum_{d\in
D_1}\mu(n/2d)+2\sum_{d\in D_2\cup \widetilde{D_3}}\mu(n/2d). $ For
$d\in \widetilde{D_3}$ we conclude that $4\mid n/2d$ and so
$\sum_{d\in \widetilde{D_3}}\mu(n/2d)=0$. Now, the formula for the
eigenvalue $\lambda_2$ becomes $\lambda_2=\sum_{d\in
D_1}\mu(n/2d)+\sum_{d\in D_1}\mu(n/2d)+2\sum_{d\in D_2}\mu(n/2d)$.
Since $n/4 \notin D$, Lemma \ref{lem:D''1=2D_2} now leads to
$$
\lambda_2=2(\sum_{d\in 2D_2}\mu(n/2d)+\sum_{d\in
D_2}\mu(n/2d))=2(\sum_{d\in D_2}\mu(n/4d)+\sum_{d\in
D_2}\mu(n/2d))=0.
$$

\noindent {\bf Case 2.} $n/2\in D.$  Suppose also that $n/4 \in D.$
This gives $D_0=2(D_1\setminus \{n/2\})$ and $D_1=2(D_2\setminus
\{n/4\})$, which follows from Corollary \ref{cor:1} and Lemma
\ref{lem:D''1=2D_2}. In this case $\lambda_2$ can be written as
follows

$$
\lambda_2=\sum_{d\in D'_1 }\mu(n/d)+\sum_{d\in
D_1\setminus\{n/2\}}\mu(n/2d)+2\sum_{d\in
D_2\setminus\{n/4\}}\mu(n/2d)+\mu(1)+2\mu(2)=0+1-2=-1.
$$

Notice that $\sum_{d\in D_0}\mu(n/d)+\sum_{d\in
D_1\setminus\{n/2\}}\mu(n/2d)+2\sum_{d\in D_2\setminus\{n/4\}}=0$
according to Case 1 of the proof.
\medskip
In both cases we conclude $\lambda_1=\lambda_2$ (according to Lemma \ref{criterion1}) and  this is a
contradiction according to Corollary \ref{cor:eql}.
\end{dok}

\bigskip

{\it Perfect quantum communication distance} (PQCD) of an arbitrary
pair of vertices $a$ and $b$ is the distance $d(a,b)$ if a perfect
state transfer exists between them. If we consider a circulant
network with identical couplings PST occurs only between vertices
$b$ and $b+n/2$ for $0\leq b\leq n/2-1$ (Theorem \ref{th:odd}). For
the integral circulant graph $\ICG_n(D)$ PQCD of $b$ and $b+n/2$ is
equal to one, if $n/2\in D$. Otherwise, we have that $n/4\in D$
(Theorem \ref{thm:n/2n/4}) and thus the path $b,b+n/4,b+n/2$ shows
that PQCD is equal to two. In both cases PQCD is independent of the
order of the graph.

\bigskip

Now, we are ready to describe the spectrum of integral circulant
graphs. The criterion for existence of PST in integral circulant
graphs that we will use in the next two theorems is given by the
following lemma.

\begin{lem}
\label{criterion2}  $\ICG_n(D)$ has PST if and only if one of the
following conditions holds
\begin{itemize}
\item[{\bf i)}] $\lambda_{2j}\in 4\N+2$ and $\lambda_{2j+1}=0$, if $n/2 \not \in D$
\item[{\bf ii)}] $\lambda_{2j}\in 4\N+1$ and $\lambda_{2j+1}=-1$, if $n/2  \in D$
\end{itemize}
for $0\leq j\leq  n/2$.
\end{lem}

\begin{dok}
In the proof we will use the same notation as in the proof of
Lemma \ref{lem:D''1=2D_2}. Suppose that $\ICG_n(D)$ has PST.

From Lemma \ref{criterion1} it follows that $\lambda_{j}=0$ or
$\lambda_j=-1$ depending on whether $n/2\not \in D$ or $n/2\in D$,
for any odd $0\leq j\leq n-1$.

\medskip

Now suppose that $n/4\in D$. According to Theorem \ref{thm:n/2n/4}
we have that $n/2\not \in D$. This implies that $\lambda_0\in 2\N$
and hence all the eigenvalues of $\ICG_n(D)$ are even for $j\in
2\N$, which follows from Lemma \ref{criterion1}. Now, we proceed
using the proof of Lemma \ref{lem:D''1=2D_2}. We have that
$$
\lambda_j=2\lambda'_{j/2}
$$
where $\lambda'_{i}$ are the eigenvalues of the integral circulant
graph $\ICG_{n_1}(D')$ where $D'=D_1\cup D_2\cup
\widetilde{D_3}\setminus\{n/2\}$ and $n_1=n/2$.

Since $n/4=n_1/2\in D$ it follows that $\lambda'_0\in 2\N+1$ which
further implies that $\lambda_{0}\in 4\N+2$. Finally, from Lemma
\ref{criterion1} we conclude that $\lambda_j\in 4\N+2$ for $j\in
2\N$.
\smallskip

If $n/2\in D$ then $\lambda_0\in 2\N+1$ and hence $\lambda_{j}\in
2\N+1$ for $j\in 2\N$. This further implies that
$\lambda_j=2\lambda'_{j/2}+1$. According to Theorem
\ref{thm:n/2n/4}, it holds that $n/4=n_1/2\not \in D$ and so
$\lambda'_0\in 2\N$. This implies that $\lambda_0\in 4\N+1$.
Finally, from Lemma \ref{criterion1} we conclude that $\lambda_j\in
4\N+1$ for $j\in 2\N$.
\medskip

If any of {\bf i)} or {\bf ii)} holds, it can easily be seen that
$\lambda_{j+1}-\lambda_j\in 4\N+2$ which implies that $\ICG_n(D)$
has PST, according to Lemma \ref{lem:pow2}.

\end{dok}

\begin{thm}
\label{thm:n/2iffn/4} Let $D$ be a set of divisors of $n$ such that
$n/2,n/4\not \in D$. Then $\ICG_n(D\cup\{n/4\})$ has PST if and only
if $\ICG_n(D\cup\{n/2\})$ has PST.
\end{thm}
\begin{dok}

Let $\lambda_j$, $\mu_j$ and $\nu_j$ be the eigenvalues of graphs
$\ICG_n(D\cup\{n/4\})$, $\ICG_n(D\cup\{n/2\})$ and $\ICG_n(D)$,
respectively. We have the following relations between these
eigenvalues: $\lambda_j=\nu_j+c(j,4)$ and $\mu_j=\nu_j+c(j,2)$.
This yields that $\mu_j=\lambda_j-c(j,4)+c(j,2)$ for $0\leq j\leq
n-1$.

By direct computation we show that

\bb t_{4,j}=\frac4{\gcd(4,j)}= \left\{ \begin{array}{rl}
4, & 2 \nmid j\\
2, & S_2(j)=1\\
1, & 4 \mid j
\end{array} \right., \quad
c(j,4)=\left\{ \begin{array}{rl}
0, & j\in 2\N+1\\
-2, & j\in 4\N+2\\
2, &  j\in 4\N
\end{array} \right..
\label{cj4} \ee

From this it follows that

\bb \mu_j= \left\{ \begin{array}{rl}
\lambda_j-1, & j\in 2\N+1\\
\lambda_j+3, & j\in 4\N+2\\
\lambda_j-1, & j\in 4\N
\end{array} \right., \quad
\ee

The following two facts can now be easily deduced: for $j\in
2\N+1$, $\lambda_j=0$ if and only if $\mu_j=-1$ and for $j\in
2\N$, $\lambda_j\in 4\N+2$ if and only if $\mu_j\in 4\N+1$. To
complete the proof it only remains to apply Lemma
\ref{criterion2}.

\end{dok}

Finally we can state the next of our main results.

\begin{thm}
\label{thm:main} $\ICG_n(D)$ has PST if and only if $n\in 4\N$,
$D^*_1=2D^*_2$, $D_0=4D^*_2$ and either $n/4\in D$ or $n/2\in D$,
where $D^*_2=D_2\setminus \{n/4\}$ and $D^*_1=D_1\setminus \{n/2\}$.
\end{thm}
\begin{dok}

\noindent ($\Rightarrow$:) This is an easy consequence of Theorem
\ref{thm:4np2}, Lemma \ref{lem:D''1=2D_2}, Corollary \ref{cor:1}
and Theorem \ref{thm:n/2n/4}.

\noindent ($\Leftarrow$:) According to Theorem \ref{thm:n/2iffn/4},
this implication is sufficient to prove for $n/4 \in D$.
Furthermore, by Theorem \ref{thm:n/2n/4} we have $n/2\not \in D$.

\smallskip

Let $0\leq j\leq n-1$ be an odd number. For $d\in D_2\cup
\widetilde{D_3}$, we conclude that $c(j,n/d)=0$, which follows from
the fact that $4\mid t_{n/d,j}$. This implies that

$$
\lambda_j=\sum_{d\in 2D_1}c(j,n/d)+\sum_{d\in
D_1}c(j,n/d)=\sum_{d\in D_1}c(j,n/2d)+c(j,n/d)=0.
$$
The last equality follows from Lemma \ref{lem:cpm}.

\smallskip

Let $0\leq j\leq n-1$ be an even number. We have

$$
\lambda_j=\sum_{d\in 2D_1}c(j,n/d)+\sum_{d\in
D_1}c(j,n/d)+\sum_{d\in
D_2\setminus\{n/4\}}c(j,n/d)+c(j,4)+\sum_{d\in
\widetilde{D_3}}c(j,n/d).
$$
From Lemma \ref{lem:cpm} and relation (\ref{cj4}) it follows that
$$
\lambda_j=2\sum_{d\in D_1}c(j,n/d)+\sum_{d\in
D_2\setminus\{n/4\}}c(j,n/d)+\sum_{d\in \widetilde{D_3}}c(j,n/d)\pm
2.
$$
Now using Lemma \ref{lem:c(j,n/d)} we obtain
$$
\aligned \lambda_j&=2\sum_{d\in
D_1=2(D_2\setminus\{n/4\})}c(j/2,n_1/d)+2\sum_{d\in
D_2\setminus\{n/4\}}c(j/2,n_1/d)+2\sum_{d\in \widetilde{D_3}}c(j/2,n_1/d)\pm 2\\
&=2\sum_{d\in
D_2\setminus\{n/4\}}(c(j/2,n_1/d)+c(j/2,n_1/2d))+2\sum_{d\in
\widetilde{D_3}}c(j/2,n_1/d)\pm 2,
\endaligned
$$
where $n1=n/2$.

Let $j\in 4\N+2$. By Lemma \ref{lem:cpm} we obtain that
$c(j/2,n_1/d)+c(j/2,n_1/2d)=0$, since $j/2\in 2\N+1$. For $d\in
\widetilde{D_3}$ we conclude that $4\mid t_{n_1/d,j/2}$ and
$c(j/2,n_1/d)=0$. Finally, we conclude that $\lambda_j=c(j,4)=-2$.

\smallskip

If $j\in 4\N$, according to Lemma \ref{lem:cpm} we show that
$$
\lambda_j=4\sum_{d\in D_2\setminus\{n/4\}}c(j/2,n_1/d)+2\sum_{d\in
\widetilde{D_3}}c(j/2,n_1/d)+ 2.
$$
Furthermore, using Lemma \ref{lem:c(j,n/d)} we have $c(j/2,n_1/d)=2
c(j/4,n_1/2d)$. In either case we conclude that $\lambda_j\in 4\N+2$
for $j\in 2\N$. Now, direct application of Lemma \ref{criterion2}
completes the proof.

\end{dok}
According to the previous result, we notice that graph
$\ICG_n(D_n\setminus\{n/2\})$ has PST for $n\in 4\N$. This class of
circulant graphs is known as cocktail-party graphs (see
\cite{sandiego1}).

\section{Conclusion}

In this paper we continue to address the question of when
circulant graphs can have perfect state transfer, and improve the
necessary condition of $n$ being divisible by $4$ given in
\cite{petkovic}. Theorem \ref{thm:main} completely characterizes
the graphs $\ICG_n (D)$ having PST. This result includes the
classes of graphs having PST found in
\cite{basic2,sandiego,petkovic}.

From the above characterization we can calculate the number of
integral circulant graphs of a given order having PST. If $n\in
8\N$, by the rule of product, the number is equal to the product of
the cardinalities of the power sets of $\{d : d \mid n,\ n/d\in
8\N\}$ and $\{d : d \mid n,\ n/d\in 8\N+4\}\setminus\{n/4\}$ times
two. If $n\in 8\N+4$ the number is equal to the cardinality of the
power set of $\{d : d \mid n,\ n/d\in 8\N+4\}\setminus\{n/4\}$ times
two. In either of cases we have two possibilities since either
$n/2\in D$ or $n/4\in D$. Thus, for a given number $n$ the number of
integral circulant graphs $\ICG_n(D)$ having PST is given by the
following formula

\bb |\ICG_n(D)|= \left\{ \begin{array}{rl}
2^{\tau(\frac n 4)}, & n\in 8\N+4\\
2^{\tau(\frac n 8)\tau(\frac n {2^{S_2(n)}})}, & n\in 8\N\\
\end{array} \right., \quad
\ee where $\tau(n)$ denotes the number of the divisors of $n$.

We can see from the formula that for some values of $n\in 8\N$ (for
example $n=96,120,144,160,168,192,\ldots$) there is a great number
of graphs having PST, while for some other values $n\in 8\N+4$,
there are only $2$ such graphs. However, the number of $\ICG_n(D)$
having PST is asymptotically equal to the number of $\ICG_n(D)$ of a
given order $n$. The last conclusion follows from Corollary 7.2
given in \cite{wasin}, where it was shown that there are at most $2^{\tau(n)-1}$ integral circulant graphs on $n$ vertices.


\bigskip

It is worth mentioning that the maximum value of the {\it perfect
quantum communication distance} (i.e. the maximal distance between
vertices where a perfect state transfer occurs) is equal to $2$ for
every $n \in 4\N$. Thus, it is still an open problem whether one can
construct a network with identical couplings in which any quantum
state can be perfectly transferred over a larger distance than
$2\log_3 n$, obtained in \cite{fizicarski,fizicarski1} for two--link
hypercubes with $n$ vertices.

 An improvement of the perfect quantum communication distance is
 made
in \cite{fizicarski} by considering fixed but different
nearest-neighbor couplings. A similar approach used on circulant
graphs (having a weighted adjacency matrix) might also enlarge the
perfect quantum communication distance. Many recent papers propose
such an approach \cite{godsil,iranci,sandiego,sandiego1}. First
results concerning characterization and finding new classes of
weighted circulant graphs are given in \cite{basic3}.
Characterization of integral circulant graphs (moreover circulant
graphs) having PST is the first step in describing the more
general class of weighted integral circulant graphs having PST.
The approach to this problem should, in our opinion, use the
interplay of graph and number theory.






\bigskip










\bigskip
\noindent {\bf Acknowledgement:} This work was supported by Research
Grant 144011 of Serbian Ministry of Science and Technological
Development. The authors are grateful to the anonymous referees
whose valuable comments resulted in improvements to this article.

\end{document}